# The Relationship between Constitutions, Socioeconomics, and the Rule of Law: A Quantitative Thermodynamic Approach

Klaus Jaffe[1], Edrey Martinez[2], Ana Cecilia Soares[2], Jose Gregorio Contreras[2], Juan C Correa[3], Antonio Canova[2][*]

1. Universidad Simón Bolívar, Caracas, Venezuela
2. Universidad Católica Andres Bello, Caracas, Venezuela
3. CESA School of Business, Bogotá, Colombia

* Corresponding author

**Abstract**

Based on what we know about the thermodynamics of synergy, we explored the relationship between countries' sociocultural order (negentropy), as estimated by their constitutions, academic development, and national indicators of their rule of law and free energy (specifically, their amount of useful work, productivity, and socioeconomic health). An analysis of the 219 indicators assessed in this study revealed strong correlations between estimates of the rule of law and the number of national academic publications with socioeconomic health indicators (i.e., GDP, the Human Development Index, and infant mortality). In contrast, correlations with the length of constitutions (the numbers of words and articles) indicate that the proliferation of legal rules hinders the rule of law and socioeconomic development, and that underdevelopment and/or insufficient enforcement of national laws fosters the proliferation of legal rules. These findings suggest that no one order favors productivity and that excessive regulation and central tutelage increase social entropy and decrease socioeconomic health.

**Introduction**

New ways are needed to consiliate[1] data, measurements, and theory in order to advance social sciences. In this study we explore such an approach, based on insights from thermodynamics of complex social systems for social systems. As an descriptive exercise, let us use the case of estimating the free energy and entropy of the projection of a solid ball by a cannon. If the ball is placed upon a heap of gunpowder and the power is ignited, it may travel a few meters. But if the ball is placed into a cannon with only a little gunpowder, it might travel long distances at high speed after the explosion of the gunpowder. The difference in free energy *Φ* (spatial reach and kinetic power) of the canon ball in both situations is enormous.

The difference of the border conditions or "order" regulating the explosion and propulsion of the ball in both situations is also enormous. One scalar of the equation is Free Energy *Φ*, the other scalar is the order, information or border conditions that are implicit in the design of the canon which we will call Negentropy *η* . This same logic can be applied to the thermodynamics of social systems. For example, more knowledge about economic activity or of the working of nature *η* are correlated with more economic growth [5, 18, 24].

Here we explore a concrete empirical example of this relationship and whether constitutions can be considered as master algorithms of algorithmically infused societies.[2] A country's constitution may impact its social and economic development. We know that the constitutions of less economically and socially developed countries change very often.[3] However, detailed studies of the relationship between the particular characteristics of such constitutions and the social and economic development of the countries they are used to govern are lacking. It was recently suggested that constitutions should be considered global or transnational documents as much as national ones,[4] but we know nothing about their impact on the socioeconomic and legal health of the countries that produce them. This study is an exploration of some of the factors relevant to this relationship, based on the thermodynamics of synergy in the evolution of complex social and economic systems.[5,6]

The rationale is as follows: Synergy emerges from synchronized reciprocal positive feedback between a network of diverse specialized actors. For this process to proceed, compatible information from different sources must synchronically coordinate the actions of the intervening agents. This results in a nonlinear increase in the useful work the system can manage. In contrast, noise produces incompatible information mixes. In terms of thermodynamics, synergy is produced by gains in free energy density (*Φ*) and information density (negentropy, *η)*. In a society, the gains in *η* reflect an increase in the successful handling of constraints by coordinated agents that emancipate the system from the environment, thereby increasing productivity, efficiency, capacity for flexibility, self-regulation, and self-control of behavior through a synchronized division of ever-more-specialized labor. Some aspects of this concept are related to the constraints of the border conditions that define a system and its information content,[7] while others are amenable to empirical research, as we show here, even in the absence of a clear conceptual framework for the fuzzy notion of negentropy.[8] We refer to *Φ* as the energy used to produce useful work in a given country and to *η* as estimates of the information and/or negentropy used to manage *Φ*. Although *η* cannot be measured directly, it can be estimated by proxy variables that relate to it, such as characteristics of the rules that govern a society or the amount of knowledge a society can manage and produce.

Several examples using empirical data illuminate the relationship between *Φ* and *η*,[5] leading to the proposition of a law of irreversible thermodynamics, which states that for synergy to occur, an increase in *Φ* requires a corresponding increase in *η*:[9]

$\Delta \Phi = k_i \cdot \Delta \eta$ (1)

where *k* is a constant dependent on condition *i*. This law only applies to open systems, because in closed systems the first law of equilibrium thermodynamics states that

$\Phi - \eta = k$ (2)

Quantitatively, *Φ* can be measured as work, free energy, or potential energy, whereas *η* can be measured as entropy, order, and/or information.[10]

Here we explore the usefulness of the proposition that in human social dynamics *η* is controlled by culture, laws, constitutions, and institutions, whereas *Φ* is reflected in the robustness of the socioeconomic systems in achieving people's material wellbeing. The rational of our study is that independent of the mechanisms and multiple causal chains of relations between the existence of a constitution and the wellbeing of a country, if the good intentions that motivated the writing of the constitutions do not achieve their goal, the constitution is faulty.

**Methods**

The constitutional texts of selected countries were sorted into 59 different descriptors. For each descriptor, the number of articles and words that regulate each title of the categories was assessed. We also computed the total numbers of articles and words in the constitutions. (Here, only the number of words in the articles are presented). The following categories were considered: fundamental rights (life, private property, and freedom); civil and political rights; economic, social, and cultural rights; centralized and decentralized political organization; territorial sovereignty; and articles that did not fall into any of the categories mentioned. We also used five descriptors from the Comparative Constitutions Project (see supporting material).

We calculated the two-way correlation between the variables extracted from the constitutional texts and 130 additional country indicators for 2018 or the closest available year (see supporting material). The most important indicators were:
- The constitutional rankings compiled by the Comparative Constitutions Project.[11]
- The per capita gross domestic product (GDP), Human Development Index (HDI), infant mortality, and the World Bank's strength of legal rights index.[12]
- The number of academic articles per capita, using country ranking document of SCImago [13].
- The size of government, legal system and property rights, judicial independence, impartial courts, protection of property rights, military interference in rule of law and politics, integrity of the legal system, legal enforcement of contracts, regulatory costs of the sale of real property, and reliability of policy, based on data provided by the Fraser Institute.[14]
- Transparency International's corruption perception values, where high scores represent lower perception of corruption.[15]
- Measurements of the rule of law as lived and perceived around the world, according to household and expert surveys administered by the World Justice Project. The measurements are based on restrictions on the powers of the government, absence of corruption, open government, fundamental rights, order and security, compliance with regulations, civil justice, criminal justice, and informal justice.[16]

- The Worldwide Governance Indicators (WGIs) based on surveys, which cover households, experts, business information providers, non-governmental organizations, and public sector organizations.[17]

**Supplementary description**

Supporting Material at: **Excel**: https://bit.ly/3nJdUZr   or  **Pdf:** https://bit.ly/3l7D7LA

**Results**

Most of the indicators used correlated somewhat with indicators for the rule of law or with characteristics of each constitution. Of the 219 indicators explored (see Supporting material), only the ones showing the strongest correlations are shown in Figure 1. The variables describing characteristics and features of a country that we considered to be related to $\Phi$ or $\eta$ clustered the countries studied in two groups. One group of countries, which included the rich ones, had high scores for features for $\eta$ such as indicators of Rule of Law, Press Freedom, Academic Productivity; and high scores for proxy indices for $\Phi$ such as the Human Development Index, GDPc and low Infant Mortality; whereas the other group had low scores for these values. The indicators also grouped into two clusters. One cluster included characteristics of constitutions, WJP, Gover and Infant Mortality; the other all the rest, including GDPc. The indicators in this first cluster were positive for poor countries and negative for prosperous ones, opposite to the indicators of prosperity such as GDPc that characterized prosperous countries.

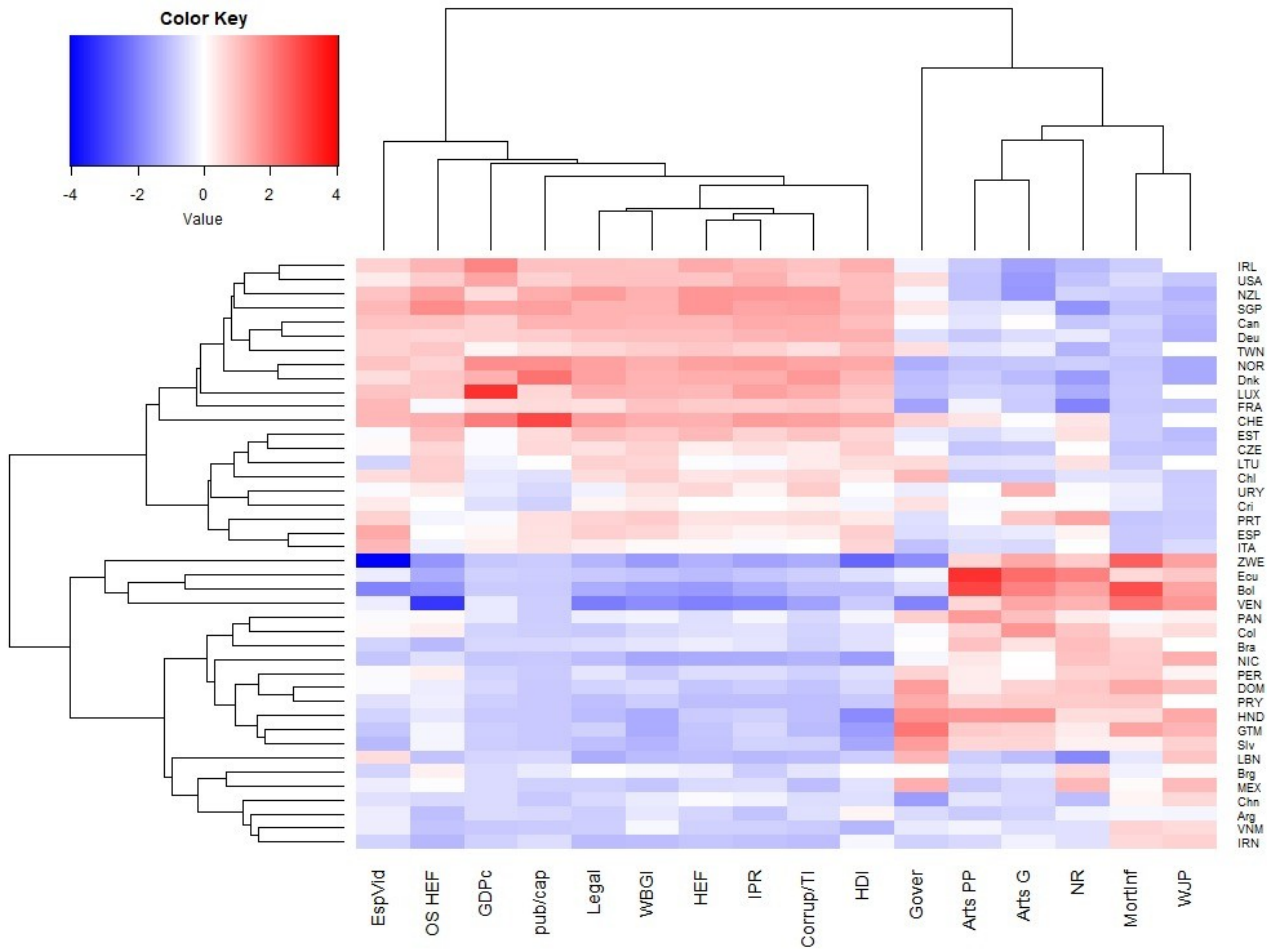

**Figure 1:** Clustering of countries according to socioeconomic variables; and clustering of these variables according to their occurrence in the countries studied. Variables are: **Espvid**: Life Expectancy; **OSHEF:** Over all Score (Heritage Foundation)**; GDPc:** Gross Domestic Product per capita; **pub/cap**: Academic articles published per capita; **Legal**: Legal System and Property Rights (Fraser Institute); **WBGI:** The worldwide Governance Indicators (RoL. Rank)**; HEF:** (Heritage Foundation) Average Rule of Law between PR, GI, JE **; IPR:** International Property Rights Index (Human Progress) ; **Corrup/TI**: Corruption Index (Transparency International); **HDI**: Human Development Index; **Gover**: Size of Government; **Arts PP:** Number of Articles for Government Prerogatives; **Arts G**: Total nr of articles in a constitution; **NR**: Nr of rights in a constitution; **Mortinf**: Infant Mortality; **WJP**: World Justice Project, ranking

Various statistical analyses showed that most variables overlapped in their predictive power as also evidenced in Figure 1. Thus, we prefer to treat these variables as interlinked, indicative of a syndrome, rather than independent and depended variables. However, as most social scientist are familiar with multiple linear regression, despite their very limited usefulness for analyzing complex systems, we present in Table 1 one example showing that Infant Mortality was better in predicting the number of articles in a constitution than other variables.

[Table 1 about here]

That is, the prosperous countries are visually recognizable in Fig 1 by showing high values in economic performance (GDPc), social performance (HDI), rule of law (Legal), etc; whereas the non-prosperous country group shared characteristics such as high infant mortality (Mortinf), week governance (Gover) and constitutions with a large number of articles (Arts G). We might use the value of these variables to define a prosperity and a poverty syndrome that help classify the countries studied into two distinct groups. This allows us to recognize variables measuring $\eta$ that are associated with positive $\Phi$. Our results show that abundance of academic publications is associated with positive $\Phi$ values (Figure 2); whereas abundance of articles in the countries constitution is associated to negative aspects of $\eta$ (Figure 3) and $\Phi$ (Figure 4).

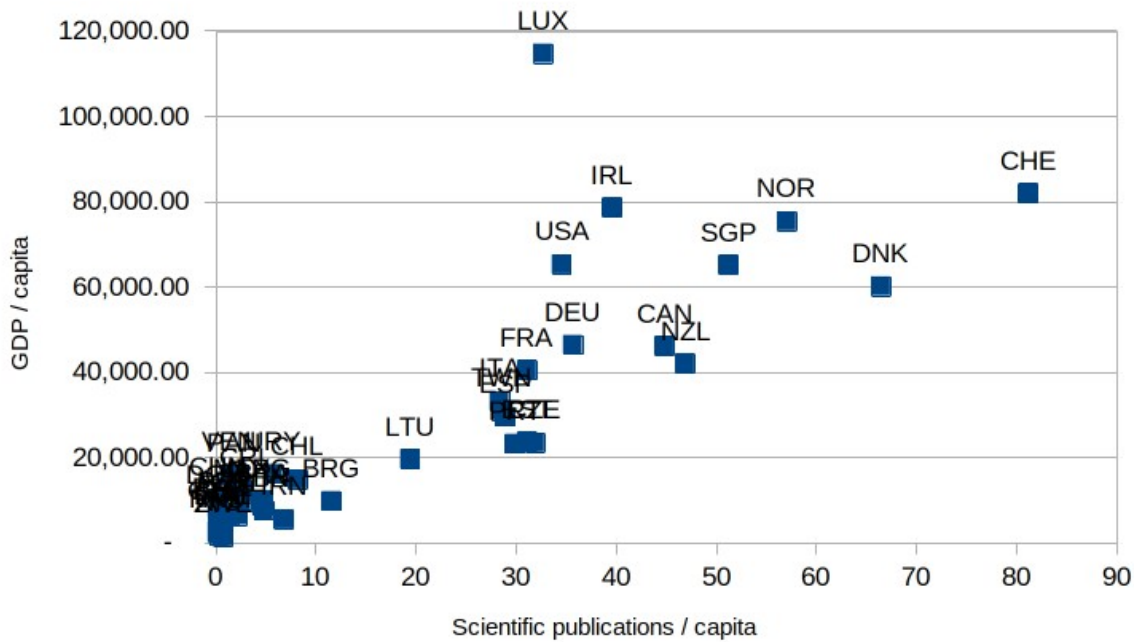

**Figure 2**. The relationship between GDP per capita (GDPc) and publications per capita (pub/cap).[12,13]

For example, the correlations that confirms a positive relationship (NOT direct causal relationships) between values for $\Phi$ and $\eta$ in Figure 2 is the strong positive relationship between publications per capita ($\eta$) and GDP per capita ($\phi$). Luxembourg, the outlier in this graph, has a high GPD not because of its production of usable goods but because it siphons taxes from its neighbors. Clearly, more information (less entropy), as reflected by the academic productivity of a country, is strongly related to its economic output.

In contrast, the indicators we calculated to characterize constitutions were correlated with indicators for $\phi$ but in the opposite direction. As shown in in Figure 3, strong perception of the Rule of Law does not correlate positively with the length of the constitution of a given country, but rather the correlation is strongly negative.

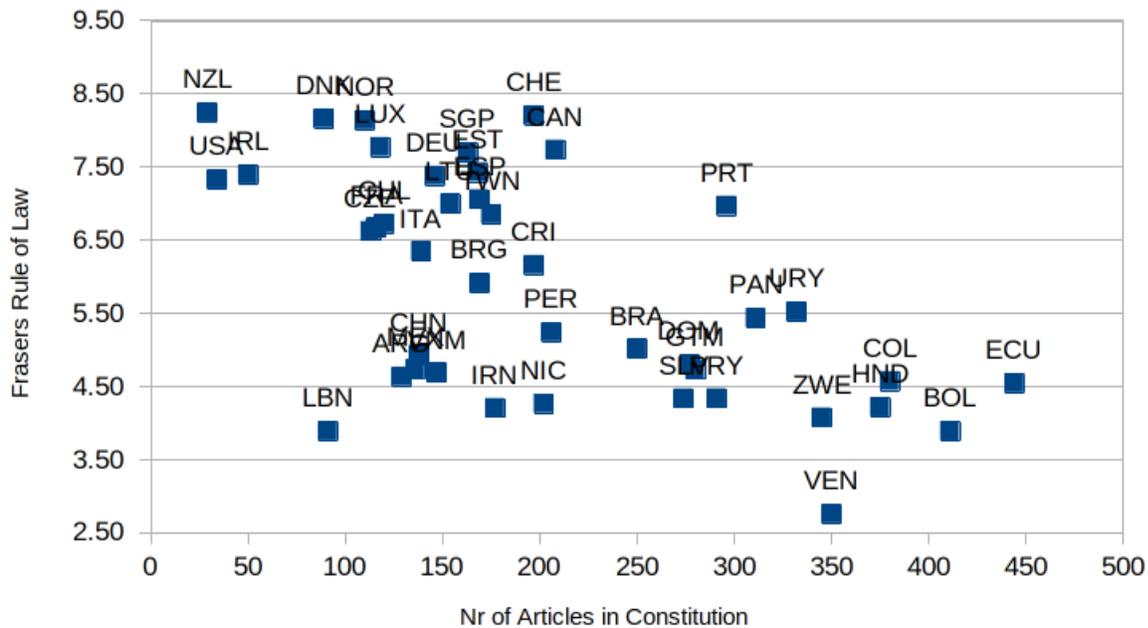

**Figure 3**. Correlation between the estimate of Rule of Law as given by Legal System and Property Rights of the Fraser Institute (Legal) and the number of articles in the constitution of that country.

A similar patter was observed with other indicators of characteristics of the constitutions. The quantitative measures related to the length of constitutions (number of articles and number of articles in a constitution addressing centralized government) and the number of rights promoted by the constitution, correlated all negatively with $\phi$, as shown in Table 2.

[Table 2 about here]

The most striking result from Table 2 is that the legal robustness of a country, as measured by intuitive subjective assessments (the indicators listed in the first four columns), is negatively correlated with the length of constitutions and the number of articles concerning centralized government. These indices, however, correlate negatively with infant mortality and positively with the HDI, a country's scientific productivity, and the length of a constitution. The length of a constitution, as measured by the numbers of articles and words it contains, gave practically identical results for all countries (see supporting material). All indicators of the strength of the rule of law used (data gathered from the World Justice Project, the Fraser Institute, Transparency International, the World Bank's Governance Indicators, and the Heritage Foundation) all had strong correlations and produced practically identical country rankings.

The indicators developed independently by the Comparative Constitutions Project (Number of Rights in Table 2) were less strongly correlated with the variables explored compared to our numerical measures of constitutions but also negatively. This was the strongest indicator of the Comparative Constitutions Project. However, even this indicator was less strongly correlated with relevant economic health indicators than the total number of articles. Interestingly, the size of government and the strength of legal rights did not correlate with any indicator of socioeconomic health or the rule of law.

The most statistically significant systemic information indicators for $\eta$ that correlated with the robustness of a country's economy and its social wellbeing ($\Phi$) were subjective indicators of the strength of the rule of law. Strong correlations between reliable objective measures for the socioeconomic health of a country, such as HDI, infant mortality, homicides, etc.; and Rule of Law and Length of Constitutions, are revealed in Table 2. Two detailed examples with correlations in opposite directions are illustrated in Figures 4 and 5 below.

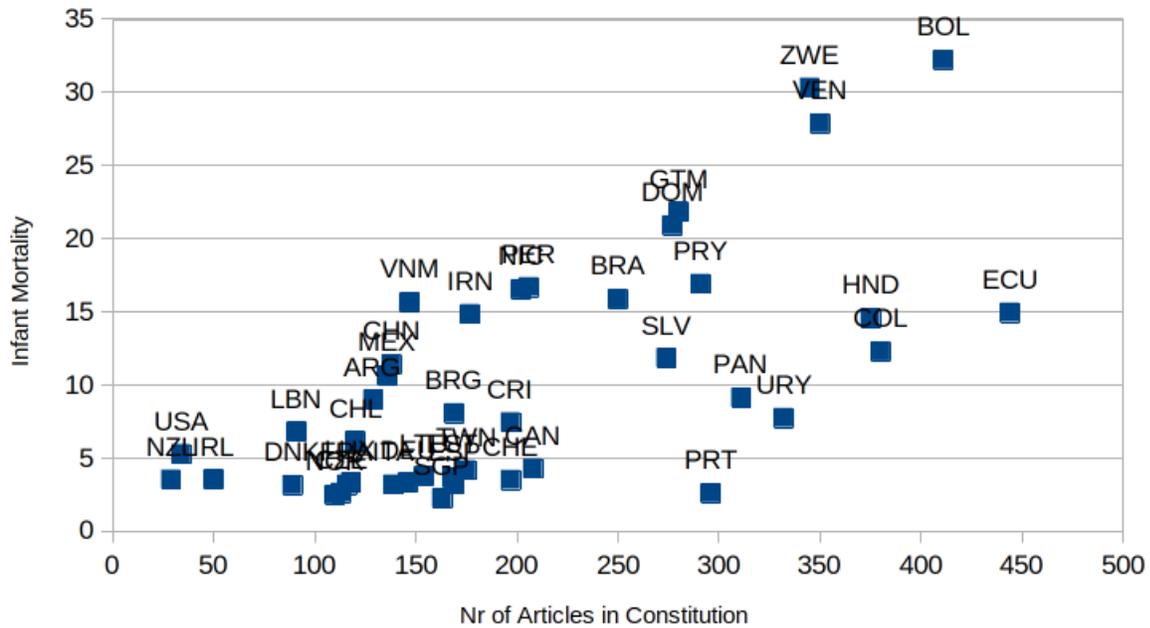

**Figure 4**. Correlation between Infant Mortality and our count of the number of articles in each country's constitution (Nr of Articles).

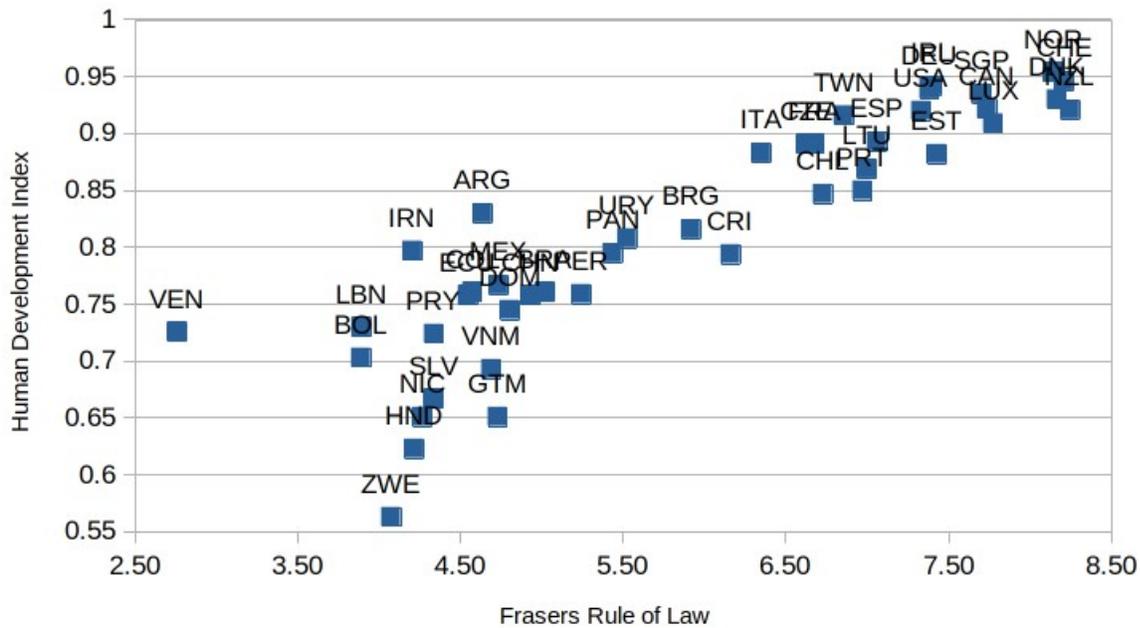

**Figure 5**. Correlations between the indicator for Legal System and Property Rights from the Fraser Institute (Legal) and the Human Development Index from the World Bank (HDI).

Figures 4 and 5 confirm the results presented in Table 1: human development is higher in countries that have a more robust legal system, but not in those that have long constitutions. Two extremes of this gradient in Figure 5 are Norway and Venezuela, both with economies dependent on oil exports.

**Discussion**

It is obvious that long constitutions do not kill infants directly. Nor that the dead of infants triggers the writing of long constitutions. Correlation is NOT causation. Our study cannot reveal the causal chain that explains our results. But the 42 countries we assessed produced results consistent with previous work on the relationship between *η* and *Φ* [18,19]. That is, the economic productivity or free energy (*Φ*), as measured by the widely accepted index of GDP per capita, is strongly correlated with the number of scientific publications (a strong quantitative index for *η*) or the amount of structured information managed by a society.[18,19]

Our results show that long constitutions do not improve the efficient enforcement of the rule of law, as indicated by the strong negative correlation between the two variables (-0.64). The unexpected negative relationship between the complexity or length of a country's constitution (*η*) and its economic robustness (*Φ*) suggests that constitutions increase rather than reduce information entropy in a society. We propose that long and verbose constitutions produce more entropy in a country's socioeconomic conditions by reducing its productivity or capacity to produce useful work (*Φ*). This view is supported by the fact that longer constitutions contained more articles concerning the functions and privileges of the state.[21] That is, longer constitutions contain more articles restraining more individual freedoms, thus hindering effective synergistic interactions between citizens.

The fact that a country's economic and social development is reflected in its constitution is an important finding that helps us understand the relationship between the rule of law, constitutions, and socioeconomic indicators such as GDP, the HDI, and infant mortality. The data suggest that the proliferation of legal rules hinders rather than encourages the rule of law and socioeconomic development. Another explanation for this correlation is that underdevelopment and/or the lack of the rule of law fosters the proliferation of legal rules.

Longer constitutions tend to increase information entropy in a society, while a higher number of academic publications is correlated with a decrease in this regard. This finding is consilient with the view that economic freedom and individual liberty of action and ideas favor the production of wealth and the reduction of corruption, whereas detailed master algorithms[1] —such as excess regulations and state tutelage [25] —hinder socioeconomic development. The exact detailed mechanisms underlying the relationship between a country having a more verbose constitution and less healthy socioeconomic achievements remain to be untangled.

But it is clear that a group of countries has achieved successful socioeconomic progress together with prolific production of scientific information and efficient social organization. Results show that lengthy constitutions are not part of their syndrome. Another group of countries has not been able to achieve this combination of factors, probably because of interference with the synergistic interactions between several sociocultural and economic factors included in the "prosperity syndrome"[22] that drive socioeconomic growth[6]. Most of these countries do have lengthy constitutions. Our results are consistent with the view that an evolutionary dynamic that includes innovation and selection favors the most adaptive information management and systems [23].

Thermodynamic considerations improve our quantitative and practical understanding of complex social dynamics, helping to rationalize our sociopolitical activities and separate useful information and order leading to useful work from noise and obstructive complications. Unfortunately, the commissions tasked with writing new constitutions in the future are unlikely to read this paper, and thus heed the advice of keeping constitutions simple and short!

**Acknowledgments:** This manuscript was submitted as a pre-print in :
https://arxiv.org/abs/2108.02094

**Table 1.** Analysis of Variance (ANOVA) for **Arts G**

|  | df | SS | MS | F | Significance |
|---|---|---|---|---|---|
| Regression | 3.000 | 204094.266 | 68031.422 | 10.627 | 0.000 |
| Residual | 38.000 | 243274.210 | 6401.953 | | |
| Total | 41.000 | 447368.476 | | | |

|  | Coefficients | Standard Error | t-Statistic | P-value |
|---|---|---|---|---|
| Intercept | 200.886 | 135.893 | 1.478 | 0.148 |
| Legal | -10.490 | 22.468 | -0.467 | 0.643 |
| MortInf | 6.707 | 2.747 | 2.441 | **0.019** |
| pubcap | -0.214 | 1.258 | -0.170 | 0.866 |

Multiple regression analysis using abbreviations as for Figure 1

**Table 2.** Pearson correlation coefficients between selected variables

|  | WJP | Transparency International | WBGI | HEF | Pub/c | HDI | Infant Mortality | GDPc |
|---|---|---|---|---|---|---|---|---|
| Better if | - | + | + | + | + | + | - | + |
| Size of Government | -0.20 | -0.20 | -0.17 | -0.26 | -0.23 | -0.23 | 0.01 | -0.25 |
| Strength of Legal Rights (WB) | -0.17 | 0.25 | 0.17 | 0.42 | 0.23 | 0.17 | -0.25 | 0.17 |
| Legal System and Property Rights (Fraser Institute) | -0.94 | 0.96 | 0.96 | 0.85 | 0.88 | 0.89 | -0.82 | 0.80 |
| WJP | – | -0.95 | -0.97 | -0.76 | -0.82 | -0.91 | 0.85 | -0.77 |
| Transparency International | -0.95 | – | 0.96 | 0.81 | 0.88 | 0.89 | -0.80 | 0.81 |
| WBGI | -0.97 | 0.96 | – | 0.81 | 0.83 | 0.92 | -0.86 | 0.76 |
| HEF | -0.76 | 0.81 | 0.81 | – | 0.68 | 0.71 | -0.78 | 0.62 |
| Number of Articles in Constitution | 0.59 | -0.56 | -0.62 | -0.54 | -0.53 | -0.63 | 0.67 | -0.52 |
| Number of Articles for Government Prerogatives | 0.55 | -0.47 | -0.54 | -0.46 | -0.41 | -0.52 | 0.64 | -0.37 |
| Number of Rights from Constitutional Ranking | 0.52 | -0.57 | -0.56 | -0.47 | -0.49 | -0.53 | 0.57 | -0.56 |

**Note**: The first five columns give the national estimates of η, including some used in previous studies,[20] and the last three columns are estimates for Φ. More variables and their cross-correlation can be found in the supporting material.

**Abbreviations**: HEF, Heritage Economic Freedom; HDI, Human Development Index; Pub/c, academic publication per capita; WB, World Bank; WBGI, World Bank Governance Indicator; WJP, World Justice Project, GDPc: Gross Domestic Product per capita

**Correlation** values > 0.5 indicate that the probability of the null hypothesis being true is p≤0.001.